\documentclass[pra,twocolumn,showpacs,preprintnumbers,amsmath,amssymb]{revtex4}
\usepackage{amsfonts}
\usepackage{amsmath}
\usepackage{amssymb}
\usepackage{graphicx}

\begin{document}

\title{Sequential Landau-Zener transitions in spin-orbit coupled systems}

\author{Xiaolong Zhang$^{1,2}$}

\author{Jiahao Huang$^{2}$}

\author{Yuexia Zhang$^{1}$}

\author{Kelin Gao$^{3}$}

\author{Chaohong Lee$^{2}$}
\altaffiliation{Corresponding author. Email: chleecn@gmail.com}

\affiliation{$^{1}$Department of Physics, Chongqing University, Chongqing 400044, China}

\affiliation{$^{2}$State Key Laboratory of Optoelectronic Materials and Technologies, School of Physics and Engineering, Sun Yat-Sen University, Guangzhou 510275, China}

\affiliation{$^{3}$State Key Laboratory of Magnetic Resonance and Atomic and Molecular Physics, Wuhan Institute of Physics and Mathematics, Chinese Academy of Sciences, Wuhan 430071, China}

\date{\today }

\begin{abstract}
We investigate the Landau-Zener (LZ) process in spin-orbit coupled systems of single or multiple two-level (spin-$\frac{1}{2}$) particles. The coupling between internal spin states and external vibrational states, a simple spin-orbit coupling (SOC), is induced by applying a spin-dependent harmonic trap. Because of the SOC, the single-particle energy-level structures are modified by the Franck-Condon (FC) effects, in which some avoided energy-level-crossings (ELCs) are almost closed and some ELCs are opened. The close of avoided ELCs and the open of ELCs result in the FC blockade and the vibrational transitions, respectively. For a given low sweeping rate, the sequential LZ transitions of ladder-like population transition can be induced by strong SOC. We derive an analytical formula for the final population which is well consistent with the numerical results. For a given strong SOC, the sequential LZ transitions are submerged in the non-adiabatic transitions if the sweeping rate is sufficiently high. Further, we study LZ transitions of multiple interacting two-level Bose particles in a spin-dependent harmonic trap. The interplay between the SOC effects and the interaction effects is explored.
\end{abstract}

\maketitle

\section{Introduction}

Landau-Zener (LZ) problem~\cite{Landau,Zener}, a well-known fundamental problem in time-dependent quantum mechanics, concerns how non-adiabatic transition appears in a two-level system driven through an avoided energy-level-crossing. According to the quantum adiabatic theorem~\cite{Born,Kato}, if the system varies infinitely slow, non-adiabatic transition will not take place and the system will always be in an eigenstate of its instantaneous Hamiltonian. The studies of LZ transition are not only of great fundamental interests~\cite{Kral,Shevchenko,Altland,Keeling,Chen}, but also of extensive applications~\cite{Makhlin,Oliver,Niskanen,Wei,LaHaye,Petta,You,Bason} in quantum state engineering, quantum interferometry and quantum computation etc. There are lots of theoretical and experimental studies of LZ transitions in systems of decoupled internal spin states and external motional states. However, to the best of our knowledge, the LZ transitions in systems of spin-orbit coupling (SOC) are still unclear. \emph{How SOC affects a LZ process?}

It has been demonstrated that strong SOC may induce Franck-Condon (FC) blockade and vibrational sidebands. The FC blockade takes place if the FC factor, which is defined as the square of the overlap integral between the vibrational wave-functions of the two involved states, is sufficiently small to be ignorable~\cite{Franck,Condon}. The FC blockade has been found in several systems, such as molecular junctions~\cite{Koch2005,Koch2006,Joachim}, nano-tube quantum dots~\cite{Leturcq,Palyi,Ohm,Ekinci}, individual neutral atoms~\cite{Forster}, and a single trapped ion~\cite{Hu}. On the other hand, nonzero FC factors may cause vibrational sidebands~\cite{Franck, Condon, Leturcq, Forster, Park, Yu, Pasupathy, Sapmaz}, that is, the population transfer or the electronic tunnelling has been shown to excite vibrational modes. Up to now, there is still no study on FC effects in the LZ process of a spin-orbit coupled particle. \emph{What signatures of FC effects will appear in a LZ process of SOC?}

In this article, we investigate the LZ process of a spin-orbit coupled spin-$\frac{1}{2}$ particle trapped by a spin-dependent potential. We explore how SOC affects energy-level structures and the LZ process. The gaps of avoided energy-level-crossings (ELCs) become narrow when the SOC becomes strong. At the same time, some ELCs are gradually opened and then closed. The appearance of FC blockade and vibrational sidebands are direct results of the close of avoided ELCs and the open of ELCs, respectively. Under sufficiently strong SOC, in contrast to the LZ transition in a system without SOC, the sequential LZ transitions of ladder-like population transition appear. However, the probabilities of the final population come from spin up and spin down are independent on SOC strength, which affects the components of the vibrational states. We find that the FC blockade corresponds to the absence of some specific population steps. Without loss of generality, we also calculate the effects come from sweeping rate. By treating the sequential LZ transitions as a sequence of conventional two-level LZ transitions and applying the conventional two-level LZ formula again and again, we obtain an analytical formula for the final populations. Based upon the current experiment techniques, it is possible to test our prediction by a nano-tube quantum dot~\cite{Palyi}, individual neutral atoms~\cite{Forster} or a single trapped ion~\cite{Hu}. Our studies provide a unique approach for exploring SOC and FC physics via LZ processes.

Further, we consider the LZ process of multiple interacting two-level Bose particles within a spin-dependent harmonic trap. Within the frame of second quantization, we derive a multi-mode two-component Bose-Hubbard model for the considered system. The particle-particle interactions play an important role for the population transition. Due to the interplay between the SOC effects and the interaction effects, the LZ transitions are dramatically different from those of single particles.

This article is outlined as follows. In Sec. II, we introduce the Hamiltonian for the LZ process of single particles in a spin-dependent harmonic trap. In Sec. III, we analyze the population dynamics of the LZ process of single-particle systems with SOC. We concentrate our analysis on the sequential population transfer starting from the lowest vibrational state $\left|\downarrow, 0_{\downarrow}\right\rangle$. This section includes four subsections. In its subsection A, we show the FC blockade and the ladder-like population transition induced by SOC. In its subsection B, we analyze the dependence of final populations on the SOC strength. In its subsection C, we show how non-adiabatic effects submerges sequential LZ transitions. In its subsection D, we address potential applications in quantum state engineering. In Sec. IV, we derive an analytical formula for the final populations. In Sec. V, we study the LZ process of interacting two-level Bose particles in a spin-dependent harmonic trap. The interplay between the SOC effects and the interaction effects is explored. In the last Sec., we briefly summarize and discuss our results.

\section{Single-particle Landau-Zener model with spin-orbit coupling}

We consider a spin-orbit coupled particle, which may be a nano-tube
quantum dot~\cite{Palyi}, individual neutral atoms~\cite{Forster} or
a single trapped ion~\cite{Hu}, in a harmonic trap. Assuming only
two internal spin states are involved, such a particle can be
regarded as a spin-$\frac{1}{2}$ particle of two spin states: $\left\vert\uparrow\right\rangle$ and $\left\vert\downarrow\right\rangle$. In a LZ process, the two spin states
are coupled by lasers with a linearly sweeping detuning. For simplicity and without loss of generality, we concentrate our studies on one-dimensional systems. The Hamiltonian reads as,
\begin{equation}
H = H_{LZ}+H_{ho}+H_{SOC},
\end{equation}
where $H_{LZ} = -\frac{\hbar\delta(t)}{2}\sigma_{z} -\frac{\hbar\Omega}{2}\sigma_{x}$ with $\delta(t)=\delta_{0}+\alpha t$ is the conventional LZ Hamiltonian, $H_{ho}=-\frac{{\hbar}^2}{2m}\frac{\partial^{2}}{\partial z^{2}}+\frac{1}{2}m\omega_{z}^{2}z^{2}$ describes the external motion and $ H_{SOC} = m\omega_{z}^{2}z_{0}z\sigma_{z}$ characterizes the SOC. Here, $\sigma_{x,z}$ are Pauli matrices, $\hbar$ is the Planck constant, $m$ is the particle mass, $\Omega$ is the Rabi frequency, $\delta(t)$ is the detuning, and $z_{0}$ is the SOC strength. Corresponding to the potential for a simple harmonic oscillator, $V^{ho}(z) = m\omega_z^2 z^2/2$, the spin-orbit coupled particle feels a spin-dependent potential $V_{\sigma}(z,t)=V_{\sigma}^{ho}(z)+U_{\sigma}(t)$ with $V_{\sigma}^{ho}(z) = \frac{1}{2}m\omega_{z}^{2}\left(z + \sigma_{z} z_{0}\right)^{2}$ and $U_{\sigma}(t) =-\frac{1}{2}m\omega_{z}^{2}z_{0}^{2}-\frac{\hbar\delta(t)}{2}\sigma_{z}$,
where $\sigma_{z} = +1$ for $\sigma=\uparrow$ and $\sigma_{z} = -1$ for $\sigma=\downarrow$. In Fig. 1, we show the schematic diagram for the LZ
process described by Hamiltonian (1).

%%%%%%%%%%%%%%%%%%%%%%%%%%%%%%%%%%%%%%%%%%%%%%%%%%%%%%%%%%%%%%%%
\begin{figure}[htb]
\includegraphics[width=1.0\columnwidth]{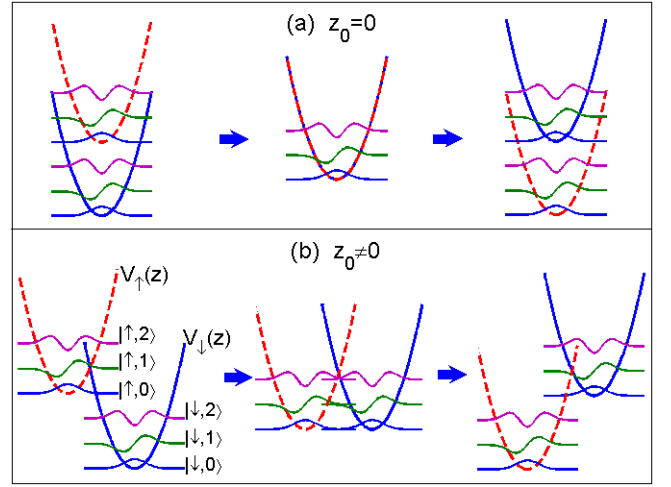}
\caption{Schematic diagram for the Landau-Zener process of a spin-$\frac{1}{2}$ particle in a harmonic trap. The red dashed curves and the blue solid curves denote the spin-dependent harmonic trap $V_{\uparrow}(z)$ and $V_{\downarrow}(z)$, respectively. The detuning is linearly swept from one far-off-resonance limit (left column) to the other far-off-resonance limit (right column) through the zero point (middle column). (a) $z_{0}=0$ for the case of no SOC and (b) $z_{0}\neq0$ for the case of SOC.}
\label{fig1}
\end{figure}
%%%%%%%%%%%%%%%%%%%%%%%%%%%%%%%%%%%%%%%%%%%%%%%%%%%%%%%%%%%%%%%%

For the system described by Hamiltonian (1), the time-evolution of its
quantum state $\left|\Psi\right\rangle =\left(\begin{array}{cc}
\psi_{\uparrow}(z,t) &  \\
\psi_{\downarrow}(z,t) &
\end{array}\right)$ obeys
\begin{eqnarray}
i\hbar\frac{\partial}{\partial t}\psi_{\uparrow} =\left[-\frac{{\hbar}^2}{2m}\frac{\partial^{2}}{\partial z^{2}} + V_{\uparrow}(z,t)\right]\psi_{\uparrow} -\frac{\hbar\Omega}{2}\psi_{\downarrow}, \\
i\hbar\frac{\partial}{\partial t}\psi_{\downarrow} =\left[-\frac{{\hbar}^2}{2m}\frac{\partial^{2}}{\partial z^{2}} + V_{\downarrow}(z,t)\right]\psi_{\downarrow}-\frac{\hbar\Omega}{2}\psi_{\uparrow},
\end{eqnarray}
with $\int dz \left(\left|\psi_{\uparrow}(z,t)\right|^2
+\left|\psi_{\downarrow}(z,t)\right|^2\right) =1$ imposed by the
normalization condition. Given the spatial eigenstates $\left\vert
n_{\sigma}\right\rangle$ for $V_{\sigma}^{ho}(z) = \frac{1}{2}m\omega_{z}^{2}\left(z + \sigma_{z} z_{0}\right)^{2}$, we have $\psi_{\sigma}(z,t) = \sum\limits_{n_{\sigma}=0}^{+\infty}C_{n_{\sigma}}(t)
\left\vert n_{\sigma}\right\rangle$ with $C_{n_{\sigma}}(t)=\left\langle
n_{\sigma} \vert \psi_{\sigma} (z,t)\right\rangle$. In the basis composed of $\left\vert \sigma, n_{\sigma}\right\rangle=\left\vert \sigma\right\rangle\otimes \left\vert n_{\sigma}\right\rangle$, the Hamiltonian (1) can be rewritten as,
\begin{eqnarray}
H&=&{\displaystyle\sum\limits_{\sigma,n_{\sigma}}} E_{n_{\sigma}}(t)\left\vert \sigma, n_{\sigma}\right\rangle \left\langle n_{\sigma},\sigma\right\vert  \notag \\
&&-\frac{\Omega}{2}{\displaystyle \sum\limits_{n_{\downarrow},n_{\uparrow}}}
\sqrt{F_{n_{\downarrow} n_{\uparrow}}} \left[ \left\vert \downarrow,
n_{\downarrow} \right\rangle \left\langle n_{\uparrow}, \uparrow\right\vert + h.c. \right].
\end{eqnarray}
Here, the FC factors $F_{n_{\downarrow} n_{\uparrow}}= F_{n_{\uparrow}
n_{\downarrow}}= \left\vert \left\langle n_{\downarrow} \right\vert \left.
n_{\uparrow}\right\rangle \right\vert ^{2}$ and $E_{n_{\sigma}}(t) =
n_{\sigma} \hbar\omega_z - \frac{1}{2}\hbar\sigma_{z}\delta(t)$ with $\sigma_{z} = +1$ for $\sigma=\uparrow$ and $\sigma_{z} = -1$ for $\sigma=\downarrow$. Comparing with Hamiltonian (1), the zero-energy point of Hamiltonian (4) is shifted to $\frac{1}{2}\left(\hbar\omega_z-m\omega_z^2z_0^2\right)$. From Hamiltonian (4), the time-evolution of amplitudes $C_{n_{\sigma}}(t)$ obeys
\begin{eqnarray}
i\hbar\frac{\partial}{\partial t}C_{n_{\uparrow}}(t)= E_{n_{\uparrow}}(t)
C_{n_{\uparrow}}(t) -\frac{\hbar\Omega}{2} {\displaystyle\sum\limits_{n_{\downarrow}}} \sqrt{F_{n_{\downarrow} n_{\uparrow}}} C_{n_{\downarrow}}(t),\\
i\hbar\frac{\partial}{\partial t}C_{n_{\downarrow}}(t)=
E_{n_{\downarrow}}(t) C_{n_{\downarrow}}(t) -\frac{\hbar\Omega}{2} {\displaystyle\sum\limits_{n_{\uparrow}}} \sqrt{F_{n_{\uparrow}
n_{\downarrow}}} C_{n_{\uparrow}}(t).
\end{eqnarray}

Obviously, if $F_{n_{\downarrow} n_{\uparrow}}=0$, there will be no
population transfer between states $\left\vert \uparrow,
n_{\uparrow}\right\rangle$ and $\left\vert \downarrow,
n_{\downarrow}\right\rangle$.

\section{Population dynamics}

In this section, we analyze the time evolutions and the population transitions in the
LZ process of SOC. By numerically integrating the coupled Schr\"{o}dinger
equations (2) and (3), we obtain $\psi_{\sigma}(z,t)$ and then calculate
both the spin populations $P_{\sigma}(t)=\int dz\left|\psi_{\sigma}(z,t)\right|^2$ and the vibrational populations $P_{\sigma n}(t) = \left|C_{n_{\sigma}}(t)\right|^2$. In our numerical
simulation, we have chosen the natural units of $m=1$, $\hbar=1$ and $\omega_z=1$. The initial state is chosen as the ground state in the negative far-off-resonance limit ($\delta<0$ and $\left|\delta\right| \gg \Omega$), that is, $\psi_{\downarrow}(z,t=0)=\left\vert \downarrow 0\right\rangle$ and $\psi_{\uparrow}(z,t=0)=0$. In the LZ processes, the Rabi frequency $\Omega$ is fixed as 0.2, the detuning is linearly swept according to $\delta=\delta_0 + \alpha t$ with the initial detuning $\delta_0=-10$ and the sweeping rate $\alpha=2.5\times10^{-4}$.

%%%%%%%%%%%%%%%%%%%%%%%%%%%%%%%%%%%%%%%%%%%%%%%%%%%%%%%%%%%%%%%%
\begin{figure}[tbp]
\includegraphics[width=1.0\columnwidth]{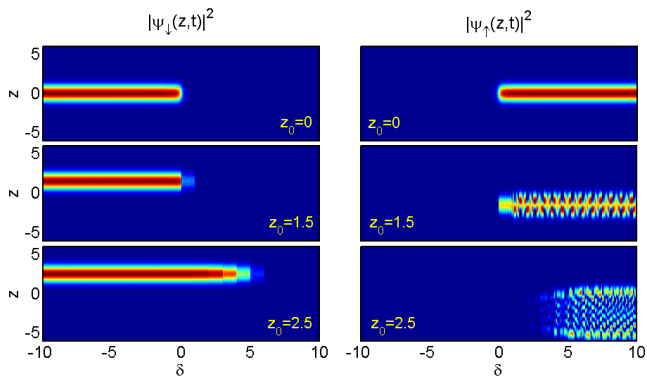}
\caption{Time evolution of probability densities in the Landau-Zener
process. The left column is $\left\vert\protect\psi_{\downarrow}(z,t)\right\vert ^{2}$ and the right column is $\left\vert \protect\psi_{\uparrow}(z,t)\right\vert ^{2}$. The top, middle and bottom rows correspond to the SOC strength $z_0=0$, 1.5 and 2.5, respectively. The other physical parameters are chosen as $\protect\delta_{0}=-10$, $\protect\alpha=2.5\times10^{-4}$ and $\Omega=0.2$.}
\label{fig2}
\end{figure}
%%%%%%%%%%%%%%%%%%%%%%%%%%%%%%%%%%%%%%%%%%%%%%%%%%%%%%%%%%%%%%%%

In Fig.~2, for different values of the SOC strength $z_0$, we show the time evolution of the probability densities $\left\vert\psi_{\sigma}(z,t)\right\vert^2$ in the LZ process. If there is no SOC, the system undergoes adiabatic evolution and the two probability densities keep in Gaussian shapes. The unchanged shapes from $\left\vert\psi_{\downarrow}(z,t)\right\vert^2$ to $\left\vert\psi_{\uparrow}(z,t)\right\vert^2$ for $z_0=0$ indicate the absence of vibrational excitations. When the SOC strength $z_0$
increases, although $\left\vert\psi_{\downarrow}(z,t)\right\vert^2$ keeps in a Gaussian shape, multi-hump structures gradually appear in $\left\vert\psi_{\uparrow}(z,t)\right\vert^2$. The appearance of multi-hump structures in $\left\vert\psi_{\uparrow}(z,t)\right\vert^2$ is a signature of vibrational excitations. In particular, the significant changes of the two probability densities sequentially take place in the vicinity of $\delta=n\hbar\omega$ (where $n$ are non-negative integers). For a larger $\alpha$, the system will undergo a non-adiabatic transition and there may be still some population in the initial state. By
controlling the sweeping rate $\alpha$, we find that it is possible
to prepare the entanglement between internal spin states and
external vibration states. The details of non-adiabatic transition
and its application in quantum state engineering will be shown in
the following subsections.

%%%%%%%%%%%%%%%%%%%%%%%%%%%%%%%%%%%%%%%%%%%%%%%%%%%%%%%%%%%%%%%%
\begin{figure}[tbp]
\includegraphics[width=1.0\columnwidth]{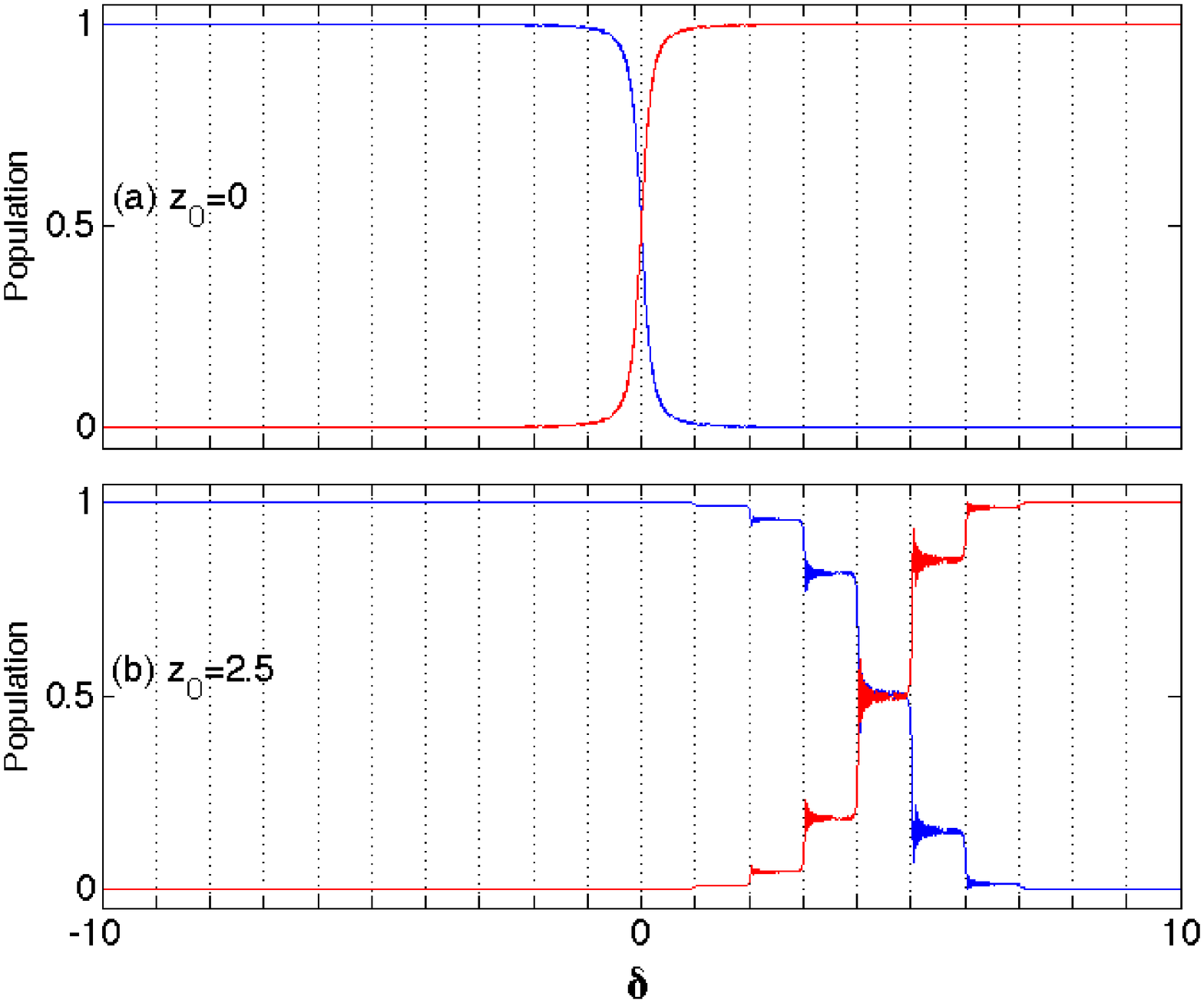}
\caption{Sequential Landau-Zener transitions induced by spin-orbit coupling.
The populations $P_{\downarrow}$ (red curves) and $P_{\uparrow}$ (blue
curves) versus the detuning $\protect\delta$ for (a) $z_{0}=0$ and (b) $z_{0}=2.5$. The sequential Landau-Zener transitions in (b) have ladder-like population transition with steps at $\protect\delta=n\hbar\protect\omega_z$.}
\label{fig3}
\end{figure}
%%%%%%%%%%%%%%%%%%%%%%%%%%%%%%%%%%%%%%%%%%%%%%%%%%%%%%%%%%%%%%%%

\subsection{Franck-Condon blockade and sequential Landau-Zener transitions}

In this subsection, we analyze how SOC induces FC blockade and sequential LZ transitions. In further, we explore the intrinsic connection between FC blockade and sequential LZ transitions.

In Fig.~3~(a), we show the population transitions corresponding to
the time evolutions in the top row of Fig.~2. Due to the absence of
SOC, the vibrational states are spin-independent and the FC factors
$F_{n_{\uparrow} n_{\downarrow}}$ are non-zero if and only if
$n_{\uparrow}=n_{\downarrow}$. Therefore, the time evolution of spin
states and vibrational states are decoupled and vibrational
excitations will not take place in the LZ process. In the
corresponding energy spectrum, the avoided ELCs only appear around
$\delta=0$ and these avoided ELCs dominate the population transfer
in the LZ process, see Fig.~4~(a). In the LZ process, as labeled by
the arrows, the system evolves along its instantaneous ground state
due to the sweeping rate $\alpha=2.5\times10^{-4}$ is sufficiently
small.

In Fig.~3~(b), we show the population transitions corresponding to the time evolutions in the bottom row of Fig.~2. Due to the strong SOC, the FC blockade and the sequential LZ transitions appear in the LZ process.
Corresponding to the significant density changes in Fig. 2, a series of
population steps appear at $\delta=n\hbar\omega$. This ladder-like
population transition is a direct signature of the sequential LZ
transitions. In addition, unlike the conventional LZ transition, there is no population step at $\delta=0$. The absence of the population step at $\delta=0$ is a result of the FC blockade between the lowest-vibrational states $\left\vert \downarrow, 0_{\downarrow}\right\rangle$ and $\left\vert \uparrow,0_{\uparrow}\right\rangle$. Based upon our numerical results for different values of $z_0$, the FC blockade appear only when $z_0$ is sufficiently large and more population steps will disappear due to the FC blockades between the lowest-vibrational state $\left\vert
\downarrow, 0_{\downarrow}\right\rangle$ and the high-vibrational states $\left\vert \uparrow, n_{\uparrow}\right\rangle$ take place for larger $z_0$. In the energy spectrum, because of the FC effects, some avoided
energy-level-crossings are almost closed due to the corresponding FC factors are very small, and some energy-level-crossings opened due to the
corresponding FC factors become non-zero, see Fig.~4~(b). Therefore, as
labeled by the arrows, the system undergoes sequential LZ transitions in
which sequential vibrational excitations accompany the ladder-like spin
population transition.

%%%%%%%%%%%%%%%%%%%%%%%%%%%%%%%%%%%%%%%%%%%%%%%%%%%%%%%%%%%%%%%%
\begin{figure}[tbp]
\includegraphics[width=1.0\columnwidth]{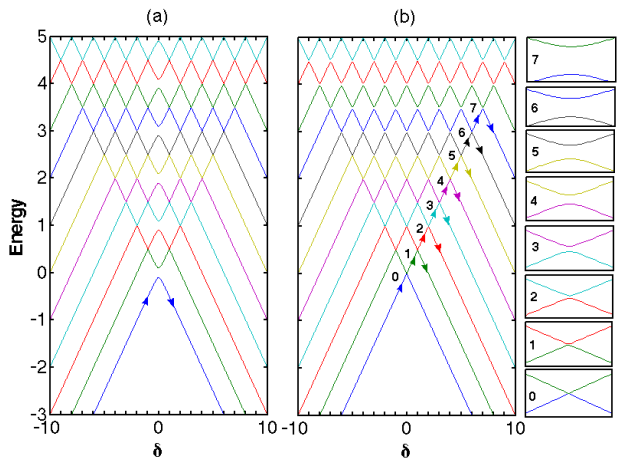}
\caption{Energy spectra for the sequential Landau-Zener transitions induced
by spin-orbit coupling. (a) and (b) correspond to (a) and (b) of Fig.~3,
respectively. The integer numbers from 0 to 7 in (b) label the (avoided)
energy-level-crossings, whose vicinities $[\protect\delta = n \pm 0.05,
\text{Energy}=\frac{n}{2} \pm 0.03]$ are magnified in the right side.}
\label{fig4}
\end{figure}
%%%%%%%%%%%%%%%%%%%%%%%%%%%%%%%%%%%%%%%%%%%%%%%%%%%%%%%%%%%%%%%%

To characterize the vibrational excitations, we analyze the population
dynamics of $\left\vert \sigma, n_{\sigma}\right\rangle$. Given $\psi_{\sigma}(z,t)$, it is easy to obtain the vibrational populations $P_{\sigma n}(t) = \left|C_{n_{\sigma}}(t)\right|^2$ with $C_{n_{\sigma}}(t)=\left\langle n_{\sigma} \vert \psi_{\sigma}
(z,t)\right\rangle$. In the absence of SOC, the vibrational populations $P_{n}(t) =P_{n}(0) = P_{\downarrow n}(t)+P_{\uparrow n}(t)$ keep unchanged. In the presence of SOC, due to the excitations from $\left\vert \downarrow,0_{\downarrow}\right\rangle$ to $\left\vert \uparrow,n_{\uparrow}\right\rangle$ around $\delta=n\hbar\omega_z$, step-like population changes appear. In other words, the population steps at $\delta=n\hbar\omega_z$ are caused by the sideband transition between $\left\vert \downarrow, 0_{\downarrow}\right\rangle$ and $\left\vert
\uparrow, n_{\uparrow}\right\rangle$.

\subsection{Final populations versus SOC strength}

In this subsection, we show how final populations depend on the SOC strength. In above, for the given slow sweeping rate, the population transitions are adiabatic when SOC is absent. For such a given slow sweeping rate, the population transitions from adiabatic to non-adiabatic when SOC becomes stronger and one may observe the FC blockade and sequential LZ transitions. In such a slow sweeping process from $\left\vert \downarrow, 0_{\downarrow}\right\rangle$, the spin population is completely inverted, but the vibrational populations sensitively depend on the SOC strength. Below, we consider the cases of a faster sweeping rate, which correspond to non-adiabatic transition even if the SOC is absent.

For a fixed sweeping rate ($\alpha=0.025$), we numerically simulate the time evolution of Eqs. (2) and (3) in LZ processes with different SOC strengthes ($z_0=0, 1.5, 2.5$). In Fig. 5, we show the spin populations $P_{\downarrow,\uparrow}$ versus the time-dependent detuning $\delta=\delta_0 + \alpha t$ for different $z_0$. The numerical results show $P_{\downarrow,\uparrow}$ are independent on $z_0$ when $\delta$ approach to positive infinity. This means that the final spin populations $P_{\downarrow,\uparrow} (t \rightarrow +\infty)$ are independent upon the SOC strength. However, the vibrational populations $P_{\sigma n} (t)$ for $\left\vert \sigma, n_{\sigma}\right\rangle$ sensitively depend on the SOC strength. As we will discussed in the following subsection, in such a fast LZ process, the step structures of sequential LZ transitions are submerged by non-adiabatic effects.

%%%%%%%%%%%%%%%%%%%%%%%%%%%%%%%%%%%%%%%%%%%%%%%%%%%%%%%%%%%%%%%%
\begin{figure}[htb]
\includegraphics[width=1.0\columnwidth]{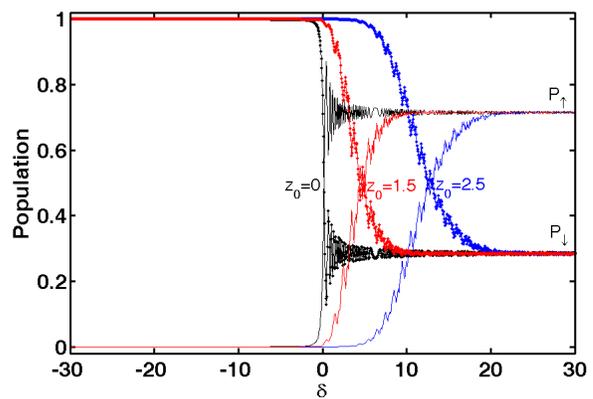}
\caption{Spin populations $P_{\downarrow,\uparrow}$ versus spin-orbit coupling strength $z_0$. The pairs of black, red and blue curves correspond to the SOC strengthes $z_{0}$=0, 1.5 and 2.5, respectively. The dotted curves denote $P_{\downarrow}$ and the solid curves denote $P_{\uparrow}$. The sweeping rate $\alpha$ is fixed at 0.025.} \label{fig.5}
\end{figure}
%%%%%%%%%%%%%%%%%%%%%%%%%%%%%%%%%%%%%%%%%%%%%%%%%%%%%%%%%%%%%%%%

\subsection{Sequential Landau-Zener transitions versus non-adiabatic effects}

In this subsection, we will show how non-adiabatic effects submerge sequential LZ transitions. For a fixed SOC strength, $z_0=1.5$ we numerically calculate the population evolution for different sweeping rates, see Fig. 6. For a small sweeping rate, $\alpha=2.5 \times 10^{-4}$, in addition to the transition from $\left\vert \downarrow, 0_{\downarrow}\right\rangle$ to $\left\vert \uparrow, n_{\uparrow}\right\rangle$, the transition from $\left\vert \downarrow, 0_{\downarrow}\right\rangle$ to $\left\vert \uparrow, 1_{\uparrow}\right\rangle$ takes place. For a moderate sweeping rate, $\alpha=2.5 \times 10^{-3}$, more population steps appear and the transition from $\left\vert \downarrow, 0_{\downarrow}\right\rangle$ to $\left\vert \uparrow, n_{\uparrow}\right\rangle$ with $n_{\uparrow}$ up to 4 takes place. However, for a large sweeping rate, $\alpha=2.5 \times 10^{-3}$, there are no significant population steps. This indicates that the non-adiabatic effects are too strong and they submerges the sequential LZ transitions.

%%%%%%%%%%%%%%%%%%%%%%%%%%%%%%%%%%%%%%%%%%%%%%%%%%%%%%%%%%%%%%%%
\begin{figure}[htb]
\includegraphics[width=1.0\columnwidth]{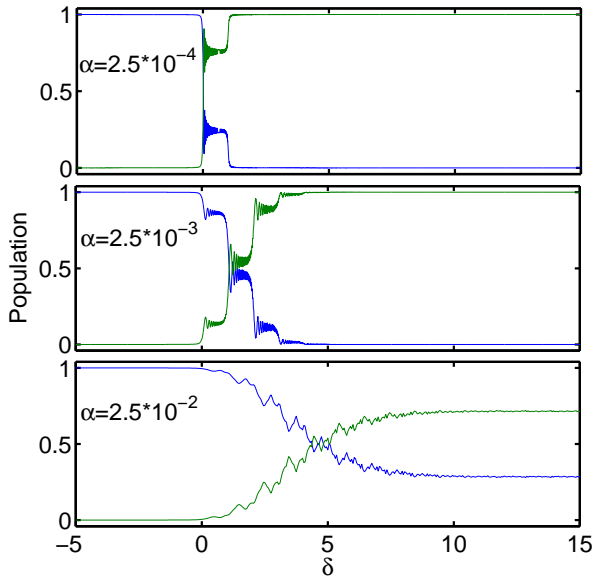}
\caption{Spin populations $P_{\downarrow,\uparrow}$ versus sweeping rate $\alpha$. The blue curves denote $P_{\downarrow}$ and the green curves denote $P_{\uparrow}$. The top, middle and bottom rows correspond to the sweeping rate $\alpha =2.5 \times 10^{-4}$, $2.5 \times 10^{-3}$ and $2.5 \times 10^{-2}$, respectively. Where the SOC strength is fixed at
$z_{0}=1.5$.} \label{fig.6}
\end{figure}
%%%%%%%%%%%%%%%%%%%%%%%%%%%%%%%%%%%%%%%%%%%%%%%%%%%%%%%%%%%%%%%%

\subsection{Potential applications in quantum state engineering}

Below, we address potential applications of the sequential LZ
transitions in preparing quantum entanglement between spin and
vibrational states. For an example, to prepare the entangled state,
$\frac{1}{\sqrt{2}}\left(\left|\downarrow,0_{\downarrow}\right\rangle +
\left|\uparrow,n_{\uparrow}\right\rangle\right)$, one can use FC
blockade in a fast or sudden sweeping to forbid the transitions from
the lowest-vibrational state
$\left|\downarrow,0_{\downarrow}\right\rangle$ to high-vibrational
states $\left|\uparrow,m_{\uparrow}\right\rangle$ with the
non-negative integer $m_{\uparrow}$ up to $n_{\uparrow}-1$. The
entanglement state can then be obtained by driving the system
through the corresponding avoided ELC with a properly small sweeping
rate $\alpha$.

In contrast to the entangled state generated by conventional dynamic
pulse, the entanglement state generated by the sequential LZ
transitions need not accurately control the pulse time. This
character enables high preparation efficiency against parameter
fluctuations. We show how to generate
$\frac{1}{\sqrt{2}}\left(\left|\downarrow,0_{\downarrow}\right\rangle
+\left|\uparrow,1_{\uparrow}\right\rangle\right)$ via sequential LZ
process, see Fig. 7. In our simulation, when the system is driven
across $\delta$=0.9, the sweeping rate is changed from
$2.5\times10^{-4}$ to $3.4\times10^{-6}$.

%%%%%%%%%%%%%%%%%%%%%%%%%%%%%%%%%%%%%%%%%%%%%%%%%%%%%%%%%%%%%%%%
\begin{figure}[htb]
\includegraphics[width=1.0\columnwidth]{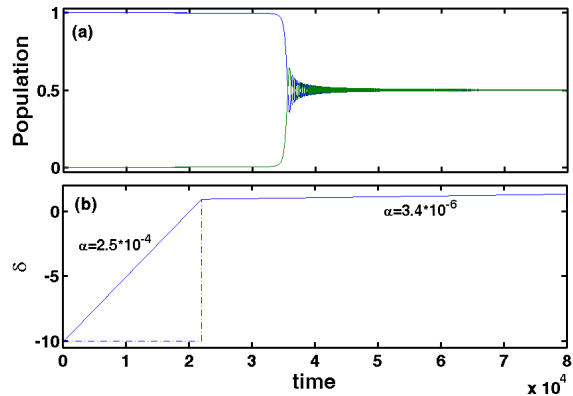}
\caption{Preparation of the entangled state $\frac{1}{\sqrt{2}}\left(\left|\downarrow,0_{\downarrow}\right\rangle +\left|\uparrow,1_{\uparrow}\right\rangle\right)$ via a two-step sweeping. In which, the system is driven from $\delta$=-10 to $\delta$=0.9 with the sweeping rate $\alpha=2.5\times10^{-4}$ and then with $\alpha=3.4\times10^{-6}$ till to $\delta=1.4$. (a) Population dynamics. The transition from $\left|\downarrow,0_{\downarrow}\right\rangle$ to $\left|\uparrow,0_{\uparrow}\right\rangle$ is forbidden by FC blockade. (b) The corresponding detuning $\delta$ versus time $t$ in the two-step sweeping process. When the time equals to $2.21\times10^{4}$ (in the natural units) we change the sweeping rate to $\alpha=3.4\times10^{-6}$. In a realistic experiment, the process swept by $\alpha$=2.5$\times10^{-4}$ can be replaced by the sudden change (dot-dashed lines). The SOC strength is at $z_{0}=2.5$.} \label{fig. 7}
\end{figure}
%%%%%%%%%%%%%%%%%%%%%%%%%%%%%%%%%%%%%%%%%%%%%%%%%%%%%%%%%%%%%%%%

Actually, the fast sweeping for enabling FC blockade can be replaced by sudden sweeping. Thus to prepare entanglement states, $\frac{1}{\sqrt{2}}\left(\left|\downarrow,0_{\downarrow}\right\rangle +\left|\uparrow,n_{\uparrow}\right\rangle\right)$, the system can be driven by a jump and a slow sweeping, see the dot-dashed lines in Fig. 7 (b). Because the FC effects blockade the transition $\left|\downarrow,0_{\downarrow}\right\rangle \leftrightarrow \left|\uparrow,0_{\uparrow}\right\rangle$ during the first sweeping process ($\delta=-10$ to $\delta=0.9$), the fast sweeping process of $\alpha=2.5\times10^{-4}$ is equivalent to the sudden change. Note that the starting point $\delta$ for generating the entangled state is insensitivity in a proper range. Whether the system stating from $\delta$=0.9 or $\delta$=0.5, it will not affect the final entangled state, $\frac{1}{\sqrt{2}}\left(\left|\downarrow,0_{\downarrow}\right\rangle +\left|\uparrow,1_{\uparrow}\right\rangle\right)$, but affect the total sweeping time. To generate the entangled state, $\frac{1}{\sqrt{2}}\left(\left|\downarrow,0_{\downarrow}\right\rangle +\left|\uparrow,2_{\uparrow}\right\rangle\right)$, the system can be driven starting from $\delta$=1.5 to $\delta$=2.5. Without loss of generality, to obtain the entangled state $\frac{1}{\sqrt{2}}\left(\left|\downarrow,0_{\downarrow}\right\rangle +\left|\uparrow,n_{\uparrow}\right\rangle\right)$, the system can be driven starting from $\delta=n-0.5$ to $\delta=n+0.5$.

\section{Analytical formula for final populations}

The sequential LZ transitions induced by SOC can be treated as a sequence of conventional two-level LZ transitions. In a conventional two-level LZ
transition of a sweeping rate $\alpha$ and the minimum gap $\Delta$ for its avoided ELC, starting from the ground state at time $t=-\infty$, the
probability of finding the system in the excited state at time $t=+\infty$ is given by the LZ formula $P_{LZ}=\exp\left(-\frac{\pi\Delta^2}{2\alpha}\right)$~\cite{Landau,Zener}. To calculate the transition probability of a two-level LZ transition in the sequence, in addition to the sweeping rate $\alpha$, one has to know the minimum gap for its avoided ELC. The minimum gap sensitively depends on the SOC strength $z_0$. By diagonalizing the effective Hamiltonian (4), the minimum gap for the avoided ELC between $\left\vert \downarrow, n_{\downarrow}\right\rangle$ and $\left\vert\uparrow, n_{\uparrow}\right\rangle$ is given as $\Delta_{n_{\uparrow}n_{\downarrow}}=\Omega \sqrt{F_{n_{\uparrow}n_{\downarrow}}}$. If there is no
SOC, $\Delta_{n_{\uparrow}n_{\downarrow}}
=\Omega\delta_{n_{\uparrow}n_{\downarrow}}$ with $\delta_{n_{\uparrow}n_{\downarrow}}$ denoting the Kronecker delta function. If the SOC is sufficiently weak, $\Delta_{00}$ is still large enough and the LZ process is similar to the conventional LZ process of no SOC. When the SOC becomes strong, the energy gaps $\Delta_{n_{\uparrow}n_{\downarrow}}$ of $n_{\uparrow} = n_{\downarrow}$ become narrow and at the same time the other energy gaps $\Delta_{n_{\uparrow}n_{\downarrow}}$ of $n_{\uparrow} \ne
n_{\downarrow}$ are gradually opened and then closed. If the SOC is
sufficiently strong, the almost vanishing $\Delta_{n_{\uparrow}n_{\downarrow}}$ may induce the blockade of the vibrational transitions $\left\vert \downarrow, n_{\downarrow}\right\rangle \Leftrightarrow
\left\vert \uparrow, n_{\uparrow}\right\rangle$ in the LZ process.

By applying the conventional two-level LZ formula to each avoided ELC, we
derive an analytical formula for the final populations. The sequential LZ
transitions can be decomposed into a sequence of conventional two-level LZ transitions: $\left\vert \downarrow, 0_{\downarrow}\right\rangle
\Leftrightarrow \left\vert \uparrow, 0_{\uparrow}\right\rangle$, $\left\vert\downarrow, 0_{\downarrow}\right\rangle \Leftrightarrow \left\vert \uparrow,1_{\uparrow}\right\rangle$, $\cdots$, $\left\vert \downarrow,0_{\downarrow}\right\rangle \Leftrightarrow \left\vert \uparrow,n_{\uparrow}\right\rangle$, $\cdots$, see Fig. 4~(b). Therefore, by applying the conventional two-level LZ formula one by one, the final populations are given as
\begin{eqnarray}
P_{\uparrow 0}=1-P_{LZ}(\Delta_{00},\alpha),  \notag \\
P_{\uparrow 1} = P_{LZ}(\Delta_{00},\alpha) (1-P_{LZ}(\Delta_{01},\alpha)),\notag \\
P_{\uparrow 2} = P_{LZ}(\Delta_{00},\alpha) P_{LZ}(\Delta_{01},\alpha)
(1-P_{LZ}(\Delta_{02},\alpha)),  \notag \\
\cdots,  \notag \\
P_{\uparrow n} =\left[\prod \limits_{n^{\prime}=0}^{n^{\prime}=n-1}
P_{LZ}(\Delta_{0n^{\prime}},\alpha)\right]\left(1-P_{LZ}(\Delta_{0n},\alpha)%
\right), \\
\cdots,  \notag \\
P_{\downarrow n}=\left\{
\begin{array}{r}
\prod\limits_{n^{\prime}=0}^{n^{\prime}=+\infty}
P_{LZ}(\Delta_{0n^{\prime}},\alpha), \quad \text{for}\quad n=0, \\
0, \quad \text{for}\quad n \ne 0.
\end{array}
\right.  \notag
\end{eqnarray}
Where, $P_{LZ}(\Delta_{0n},\alpha)=\exp\left(-\frac{\pi \Delta_{0n}^2}{2\alpha}\right)$ is given by the conventional two-level LZ formula~\cite{Landau,Zener}. By regarding our problem as a multi-state LZ problem, the survival probability $P_{\downarrow 0}$ is consistent with the one derived by the S-matrix theory~\cite{Shytov} or the perturbation analysis~\cite{Volkov} when the coupling strength is smaller than the level difference of the vibrational states.

%%%%%%%%%%%%%%%%%%%%%%%%%%%%%%%%%%%%%%%%%%%%%%%%%%%%%
\begin{table}[h]
\caption{Comparison between numerical and analytical results of final
populations. The first column is the SOC strength $z_0$, the second column is the FC factor, the third column is the final population obtained from numerical integration, and the last column is the final population given by the generalized LZ formula (7). The rows of $z_0=0$ and $2.5$ correspond to (a) and (b) of Fig. 3, respectively.}
\label{tab:final-populations}
\begin{center}
\resizebox{0.48\textwidth}{!}{
\begin{tabular}{llll}\hline\hline SOC & FC Factor     &
Numerical & Analytical\\ \hline $z_0=0$   & $F_{00}=1$       &
$P_{\uparrow 0}=0.9998$  & $P_{\uparrow 0}=1.0000$\\\hline
$z_0=2.5$ & $F_{00}=3.7\times10^{-6}$  & $P_{\uparrow 0}=0.0005$  & $P_{\uparrow 0}=0.0005$\\
          & $F_{01}=4.7\times10^{-5}$  & $P_{\uparrow 1}=0.0059$  & $P_{\uparrow 1}=0.0058$\\
          & $F_{02}=0.0003$  & $P_{\uparrow 2}=0.0367$  & $P_{\uparrow 2}=0.0357$\\
          & $F_{03}=0.0012$  & $P_{\uparrow 3}=0.1383$  & $P_{\uparrow 3}=0.1355$\\
          & $F_{04}=0.0038$  & $P_{\uparrow 4}=0.3152$  & $P_{\uparrow 4}=0.3117$\\
          & $F_{05}=0.0095$  & $P_{\uparrow 5}=0.3535$  & $P_{\uparrow 5}=0.3556$\\
          & $F_{06}=0.0197$  & $P_{\uparrow 6}=0.1378$  & $P_{\uparrow 6}=0.1423$\\
          & $F_{07}=0.0353$  & $P_{\uparrow 7}=0.0121$  & $P_{\uparrow 7}=0.0128$\\ \hline\hline
\end{tabular}}
\end{center}
\end{table}

%%%%%%%%%%%%%%%%%%%%%%%%%%%%%%%%%%%%%%%%%%%%%%%%%%%%%%

The final populations given by the analytical formula (7) are well
consistent with our numerical results. In Table I, for different SOC
strengthes $z_0$, we compare the final populations estimated by the
analytical formula (7) with the corresponding ones obtained by numerical
integration. For the case of $z_0=0$ shown in Fig.~3~(a), the vibrational
number $n$ is exactly truncated at $0$ because there is no SOC. For the case of $z_0=2.5$ shown in Fig.~3~(b), the vibrational number $n$ can be
approximately truncated at $7$. In particular, the FC factor $F_{00}$ is in order of $10^{-6}$ and the final population $P_{\uparrow 0}$ is in order of $10^{-4}$. As the initial state is $\left\vert \downarrow,
0_{\downarrow}\right\rangle$, such a small $P_{\uparrow 0}$ is a signature of the FC blockade between $\left\vert \downarrow,
0_{\downarrow}\right\rangle$ and $\left\vert \uparrow,
0_{\uparrow}\right\rangle$. For a stronger SOC, there may appear more FC
blockades between $\left\vert \downarrow, 0_{\downarrow}\right\rangle$ and $\left\vert \uparrow, n_{\uparrow}\right\rangle$ with $n_{\uparrow}$ up to a larger integer number. The numerical and analytical results clearly show that the largest absolute population difference is less than $0.005$ and the largest relative difference is less than $5.5\%$. This means that the analytical formula (7) is a very good estimation for the final populations.

\section{Interplay between spin-orbit coupling and particle-particle interaction}

In this section, we investigate the LZ process of multiple interacting two-level Bose particles, which are trapped in a spin dependent harmonic potential. We give a multi-mode two-component Bose-Hubbard model for this system and explore the interplay between the SOC effects and the interaction effects.

In quantum field theory, by using second quantization, the system of Bose condensed two-level (quasi spin-$\frac{1}{2}$) atoms can be treated as a two-component Bose field. Therefore, the system is described by the many-body Hamiltonian~\cite{Griffin,Pethick,Chao}
\begin{eqnarray}
H&=&H_{\downarrow}+H_{\uparrow}+H_{\downarrow\uparrow},
\end{eqnarray}
with the single-component Hamiltonian for the particles in spin
down state $\left|\downarrow\right\rangle$,
\begin{eqnarray}
&H_{\downarrow}& = \int\hat{\psi}_{\downarrow}^{\dagger}(\textbf{r},t) [-\frac{\hbar^2\nabla^2}{2m} +V_{\downarrow}(\textbf{r},t)] \hat{\psi}_{\downarrow}(\textbf{r},t)d\textbf{r}\nonumber\\
&&+\frac{g_{\downarrow\downarrow}}{2} \int\hat{\psi}_{\downarrow}^{\dagger}(\textbf{r},t) \hat{\psi}_{\downarrow}^{\dagger}(\textbf{r},t) \hat{\psi}_{\downarrow}(\textbf{r},t) \hat{\psi}_{\downarrow}(\textbf{r},t)d\textbf{r}\nonumber\\
&&-\frac{\delta(t)}{2} \int\hat{\psi}_{\downarrow}^{\dagger}(\textbf{r},t) \hat{\psi}_{\downarrow}(\textbf{r},t)d\textbf{r},\nonumber
\end{eqnarray}
the other single-component Hamiltonian for the particles in spin up state $\left|\uparrow\right\rangle$,
\begin{eqnarray}
&H_{\uparrow}& =\int\hat{\psi}_{\uparrow}^{\dagger}(\textbf{r},t) [-\frac{\hbar^2\nabla^2}{2m} +V_{\uparrow}(\textbf{r},t)] \hat{\psi}_{\uparrow}(\textbf{r},t)d\textbf{r}\nonumber\\
&&+\frac{g_{\uparrow\uparrow}}{2} \int\hat{\psi}_{\uparrow}^{\dagger}(\textbf{r},t) \hat{\psi}_{\uparrow}^{\dagger}(\textbf{r},t) \hat{\psi}_{\uparrow}(\textbf{r},t) \hat{\psi}_{\uparrow}(\textbf{r},t)d\textbf{r}\nonumber\\
&&+\frac{\delta(t)}{2} \int\hat{\psi}_{\uparrow}^{\dagger}(\textbf{r},t) \hat{\psi}_{\uparrow}(\textbf{r},t)d\textbf{r},\nonumber
\end{eqnarray}
and the inter-component interaction and the linear coupling between particles in different spin states,
\begin{eqnarray}
&H_{\downarrow\uparrow}& =g_{\downarrow\uparrow} \int\hat{\psi}_{\uparrow}^{\dagger}(\textbf{r},t) \hat{\psi}_{\downarrow}^{\dagger}(\textbf{r},t) \hat{\psi}_{\downarrow}(\textbf{r},t) \hat{\psi}_{\uparrow}(\textbf{r},t)d\textbf{r}\nonumber\\
&&-\frac{\hbar\Omega}{2} \int[\hat{\psi}_{\downarrow}^{\dagger}(\textbf{r},t) \hat{\psi}_{\uparrow}(\textbf{r},t) +\hat{\psi}_{\uparrow}^{\dagger}(\textbf{r},t) \hat{\psi}_{\downarrow}(\textbf{r},t)]d\textbf{r}.\nonumber
\end{eqnarray}
Here, $m$ is the single-particle mass, $g_{\sigma\sigma'}=4\pi\hbar^2a_{\sigma\sigma'}/m$ with $a_{\sigma\sigma'}$ denoting the s-wave scattering length between the particles in spin states $\left\vert \sigma\right\rangle$ and $\left\vert \sigma'\right\rangle$, $V_{\sigma}(\textbf{r},t)$ is the spin-dependent harmonic potential, $\delta(t)$ is the detuning and $\Omega$ is the Rabi frequency. The symbols $\hat{\psi}_{\sigma}^{\dagger}(\textbf{r},t)$ and $\hat{\psi}_{\sigma}(\textbf{r},t)$ are Bose creation and annihilation operators for particles in spin state $\sigma$, respectively.

We now show the derivation of the multi-mode two-component Bose-Hubbard model with the multi-mode expansion. Denoting the $l-$th single-particle eigenstate for $V_{\sigma}(\textbf{r},t)$ as $\phi_{\sigma}^{l}(\textbf{r})$, the atomic fields can be expanded as
\begin{equation}
\hat{\psi}_{\sigma}(\textbf{r},t) =\sum_{l}\hat{b}_{\sigma}^{l} \phi_{\sigma}^{l}(\textbf{r}),\nonumber
\end{equation}
with $\hat{b}_{\sigma}^{l}$ being the annihilation operators of the atoms in the $l$-th eigen-mode and the spin state $\left|\sigma\right\rangle$. By integrating all spatial degrees of freedom, we have
\begin{eqnarray}
&H_{\downarrow}&=\sum_{l}\epsilon_{\downarrow}^l \hat{n}_{\downarrow}^l +\sum_{l_{1}l_{2}l_{3}l_{4}} \frac{U_{\downarrow\downarrow}^{l_{1}l_{2}l_{3}l_{4}}}{2} \hat{b}_{\downarrow}^{l_{1}\dagger} \hat{b}_{\downarrow}^{l_{2}\dagger} \hat{b}_{\downarrow}^{l_{3}} \hat{b}_{\downarrow}^{l_{4}},
\end{eqnarray}
\begin{eqnarray}
&H_{\uparrow}&=\sum_{l} \epsilon_{\uparrow}^l \hat{n}_{\uparrow}^l +\sum_{l_{1}l_{2}l_{3}l_{4}} \frac{U_{\uparrow\uparrow}^{l_{1}l_{2}l_{3}l_{4}}}{2} \hat{b}_{\uparrow}^{l_{1}\dagger} \hat{b}_{\uparrow}^{l_{2}\dagger} \hat{b}_{\uparrow}^{l_{3}} \hat{b}_{\uparrow}^{l_{4}},
\end{eqnarray}
\begin{eqnarray}
H_{\downarrow\uparrow} &=&-\sum_{ll'}J_{ll'} (\hat{b}_{\downarrow}^{l\dagger} \hat{b}_{\uparrow}^{l'} +\hat{b}_{\uparrow}^{l'\dagger} \hat{b}_{\downarrow}^{l})\nonumber\\
&&+\sum_{l_{1}l_{2}l_{3}l_{4}} U_{\downarrow\uparrow}^{l_{1}l_{2}l_{3}l_{4}} \hat{b}_{\uparrow}^{l_{1}\dagger} \hat{b}_{\downarrow}^{l_{2}\dagger} \hat{b}_{\downarrow}^{l_{3}} \hat{b}_{\uparrow}^{l_{4}},
\end{eqnarray}
with the tunneling strength,
\begin{eqnarray}
&J_{ll'}&=\frac{\hbar\Omega}{2} \int\phi_{\downarrow}^{l*}(\textbf{r}) \phi_{\uparrow}^{l'}(\textbf{r}) d\textbf{r},\nonumber
\end{eqnarray}
the single-particle energy for the spin-down particle,
\begin{eqnarray}
&\epsilon_{\downarrow}^l &=\int[\phi_{\downarrow}^{l*}(\textbf{r}) (-\frac{\hbar^2\nabla^2}{2m} +V_{\downarrow} -\frac{\delta}{2})\phi_{\downarrow}^{l}(\textbf{r})] d\textbf{r},\nonumber
\end{eqnarray}
the single-particle energy for the spin-up particle,
\begin{eqnarray}
&\epsilon_{\uparrow}^l& =\int[\phi_{\uparrow}^{l*}(\textbf{r}) (-\frac{\hbar^2\nabla^2}{2m} +V_{\uparrow} +\frac{\delta}{2}) \phi_{\uparrow}^{l}(\textbf{r})]d\textbf{r},\nonumber
\end{eqnarray}
the intra-component interaction,
\begin{eqnarray}
&U_{\sigma\sigma}^{l_{1}l_{2}l_{3}l_{4}}& =g_{\sigma\sigma} \int\phi_{\sigma}^{l_{1}*} \phi_{\sigma}^{l_{2}*}\phi_{\sigma}^{l_{3}} \phi_{\sigma}^{l_{4}}d\textbf{r} =g_{\sigma\sigma} Y_{\sigma\sigma}^{l_{1}l_{2}l_{3}l_{4}},\nonumber
\end{eqnarray}
and the inter-component interaction,
\begin{eqnarray}
&U_{\downarrow\uparrow}^{l_{1}l_{2}l_{3}l_{4}}& =g_{\downarrow\uparrow} \int\phi_{\uparrow}^{l_{1}*} \phi_{\downarrow}^{l_{2}*} \phi_{\downarrow}^{l_{3}} \phi_{\uparrow}^{l_{4}}d\textbf{r} =g_{\downarrow\uparrow} Y_{\downarrow\uparrow}^{l_{1}l_{2}l_{3}l_{4}}.\nonumber
\end{eqnarray}
Here, $Y_{\sigma\sigma'}^{l_{1}l_{2}l_{3}l_{4}} =\int\phi_{\sigma}^{l_{1}*} \phi_{\sigma'}^{l_{2}*}\phi_{\sigma'}^{l_{3}} \phi_{\sigma}^{l_{4}}d\textbf{r}$ are the overlap integrals for two-body interactions. In our calculation, the values of all $U_{\sigma\sigma'}^{l_{1}l_{2}l_{3}l_{4}}$ are obtained by the product of interaction strength $g_{\sigma\sigma'}$ and the overlap integral $Y_{\sigma\sigma'}^{l_{1}l_{2}l_{3}l_{4}}$. In Table II, we list the values of $Y_{\sigma\sigma'}^{l_{1}l_{2}l_{3}l_{4}}$ involving $l=0$ and/or $1$ for a system of the SOC strength $z_{0}=1.0$.

%%%%%%%%%%%%%%%%%%%%%%%%%%%%%%%%%%%%%%%%%%%%%%%%%%%%%
\begin{table}[htb]
\caption{The overlap integral for a system of the SOC strength $z_{0}=1.0$.}
\label{tab:overlap integral}
\begin{center}
\resizebox{0.40\textwidth}{!}{
\begin{tabular}{lllll}\hline\hline $Y_{\sigma\sigma}^{l_{1}l_{2}l_{3}l_{4}}$
& $Y_{\sigma\sigma}^{0000}=0.3989$ & $Y_{\sigma\sigma}^{0100}=0.0000$\\
& $Y_{\sigma\sigma}^{1000}=0.0000$ & $Y_{\sigma\sigma}^{1100}=0.1995$\\
& $Y_{\sigma\sigma}^{0010}=0.0000$ & $Y_{\sigma\sigma}^{0110}=0.1995$\\
& $Y_{\sigma\sigma}^{1010}=0.1995$ & $Y_{\sigma\sigma}^{1110}=0.0000$\\
& $Y_{\sigma\sigma}^{0001}=0.0000$ & $Y_{\sigma\sigma}^{0101}=0.1995$\\
& $Y_{\sigma\sigma}^{1001}=0.1995$ & $Y_{\sigma\sigma}^{1101}=0.0000$\\
& $Y_{\sigma\sigma}^{0011}=0.1995$ & $Y_{\sigma\sigma}^{0111}=0.0000$\\
& $Y_{\sigma\sigma}^{1011}=0.0000$ & $Y_{\sigma\sigma}^{1111}=0.2992$\\
\hline $Y_{\downarrow\uparrow}^{l_{1}l_{2}l_{3}l_{4}}$
& $Y_{\sigma\sigma'}^{0000}=0.0540$ & $Y_{\sigma\sigma'}^{0100}=-0.0764$\\
& $Y_{\sigma\sigma'}^{1000}=0.0764$ & $Y_{\sigma\sigma'}^{1100}=-0.0810$\\
& $Y_{\sigma\sigma'}^{0010}=-0.0764$ & $Y_{\sigma\sigma'}^{0110}=0.1350$\\
& $Y_{\sigma\sigma'}^{1010}=-0.0810$ & $Y_{\sigma\sigma'}^{1110}=0.1145$\\
& $Y_{\sigma\sigma'}^{0001}=0.0764$ & $Y_{\sigma\sigma'}^{0101}=-0.0810$\\
& $Y_{\sigma\sigma'}^{1001}=0.1350$ & $Y_{\sigma\sigma'}^{1101}=-0.1145$\\
& $Y_{\sigma\sigma'}^{0011}=-0.0810$ & $Y_{\sigma\sigma'}^{0111}=0.1145$\\
& $Y_{\sigma\sigma'}^{1011}=-0.1145$ & $Y_{\sigma\sigma'}^{1111}=0.1485$\\
\hline\hline
\end{tabular}}
\end{center}
\end{table}
%%%%%%%%%%%%%%%%%%%%%%%%%%%%%%%%%%%%%%%%%%%%%%%%%%%%%%

If $V_{\sigma}(\textbf{r},t)$ is a spin-independent potential and the mode labels $(l_{1}, l_{2}, l_{3}, l_{4})$ have the same value, the above multi-mode many-body Hamiltonian becomes a conventional Bose-Josephson junction~\cite{Leggett, Milburn, Steel, Cirac, Kohler, Hong, Frazer, Lee}. In the LZ process of a Bose-Josephson junction, the theoretical prediction of interaction blockade has been demonstrated in laboratories~\cite{Capelle, Kivshar, Cheinet}.

In the system of no SOC, due to the orthogonality between different vibrational eigenstates ($J_{ll'}=\frac{\hbar\Omega}{2} \delta_{l,l'}$), the tunneling between states of different $l$ is inhibited. Unlike to the systems of no SOC, in the system of SOC, the tunneling between states of different $l$ may appear due to their significant FC factor and even the tunneling between states of same $l$ may vanish due to the FC blockade.

In the single-particle system, which has been discussed in previous sections, the population transfer between states of different $l$ should be assisted by the SOC term. However, in the system of multiple interacting particles, the interaction terms of $U_{\sigma\sigma'}^{l_{1}l_{2}l_{3}l_{4}}$ allow that a spin-$\sigma$ atom and a spin-$\sigma'$ atom change their vibrational states from $\left|l_{3}, l_{4}\right\rangle$ to $\left|l_{2}, l_{1}\right\rangle$ in the collision. Therefore, the interplay between the SOC and the inter-particle interaction makes the LZ process more complex. Below, we concentrate our discussion on the interplay between the FC blockade induced by SOC and the interaction blockade induced by the two-body interactions.

To understand the LZ transitions between different instantaneous eigenstates, we analyze the energy spectra of the multi-mode two-component Bose-Hubbard Hamiltonian obeying Eqs.~(8-11). In our calculation, the diagonalization is implemented by using the Fock bases $\left\{\Pi_{\sigma=\{\downarrow, \uparrow\}} \Pi_{l=\{0,1,2,\cdots\}} \otimes \left|n_{\sigma}^{l}\right\rangle\right\}$, in which $n_{\sigma}^{l}=\hat{b}_{\sigma}^{l\dag}\hat{b}_{\sigma}^{l}$ denotes the number of atoms in the spin state $\left|\sigma\right\rangle$ and the vibrational state $\left|l\right\rangle$. Obviously, the vibrational number $l$ have infinitely possible values. However, it can be truncated at a sufficiently large number in the LZ process from the lowest vibrational states. As an example, we consider the case of two atoms, i.e., the total atomic number $N=\sum_{\sigma=\{\downarrow, \uparrow\}} \sum_{l=\{0,1,2,\cdots\}} n_{\sigma}^{l} =2$. Therefore, the bases include $\left|2_{\downarrow}^{l}\right\rangle$ (for arbitrary $l$), $\left|2_{\uparrow}^{l}\right\rangle$ (for arbitrary $l$), $\left|1_{\downarrow}^{l}, 1_{\downarrow}^{l'}\right\rangle$ (for $l \leq l'$), $\left|1_{\uparrow}^{l}, 1_{\uparrow}^{l'}\right\rangle$ (for $l \leq l'$) and $\left|1_{\downarrow}^{l}, 1_{\uparrow}^{l'}\right\rangle$ (for arbitrary $l$ and $l'$) with $l=\{0,1,2,\cdots\}$ and $l'=\{0,1,2,\cdots\}$. Because we are only interested in the LZ process involving low-energy levels, the vibrational number $l$ is truncated at $l=2$ in our numerical calculation.

In Fig.~8, we show the energy spectra for a two-particle system with no inter-particle interactions. In the system without SOC, an avoided ELC involving three lowest energy levels appears at $\delta=0$, see Fig.~8~(a). Therefore, in the LZ process from the ground state $\left|2_{\downarrow}^{0}\right\rangle$, if the sweeping is sufficiently slow, the system will evolves  adiabatically into the final state $\left|2_{\uparrow}^{0}\right\rangle$. The state at the unbiased point $\delta=0$ is a superposition state of $\left|2_{\downarrow}^{0}\right\rangle$, $\left|1_{\downarrow}^{0}, 1_{\uparrow}^{0}\right\rangle$ and $\left|2_{\downarrow}^{0}\right\rangle$. Similarly, if there is SOC, the FC blockade appears in the LZ process, see Fig.~8~(b). The first significant avoided ELC is move to the vicinity of $\delta=2$, which is marked by a pair of arrows. In the LZ process from the ground state $\left|2_{\downarrow}^{0}\right\rangle$, the complete population transfer to the final state $\left|2_{\downarrow}^{2}\right\rangle$ can be achieved by choosing a suitable sweeping rate. In such a LZ process, the population transiting from $\left|2_{\downarrow}^{0}\right\rangle$ to $\left|2_{\uparrow}^{0}\right\rangle$, $\left|1_{\downarrow}^{0}, 1_{\uparrow}^{0}\right\rangle$, $\left|1_{\uparrow}^{0}, 1_{\uparrow}^{1}\right\rangle$, $\left|2_{\uparrow}^{1}\right\rangle$, $\left|1_{\downarrow}^{0}, 1_{\uparrow}^{1}\right\rangle$, $\left|1_{\downarrow}^{1}, 1_{\uparrow}^{0}\right\rangle$ and $\left|1_{\uparrow}^{1}, 1_{\uparrow}^{2}\right\rangle$ are blockaded by their small FC factors.

%%%%%%%%%%%%%%%%%%%%%%%%%%%%%%%%%%%%%%%%%%%%%%%%%%%%%%%%%%%%%%%%
\begin{figure}[htb]
\includegraphics[width=1.0\columnwidth]{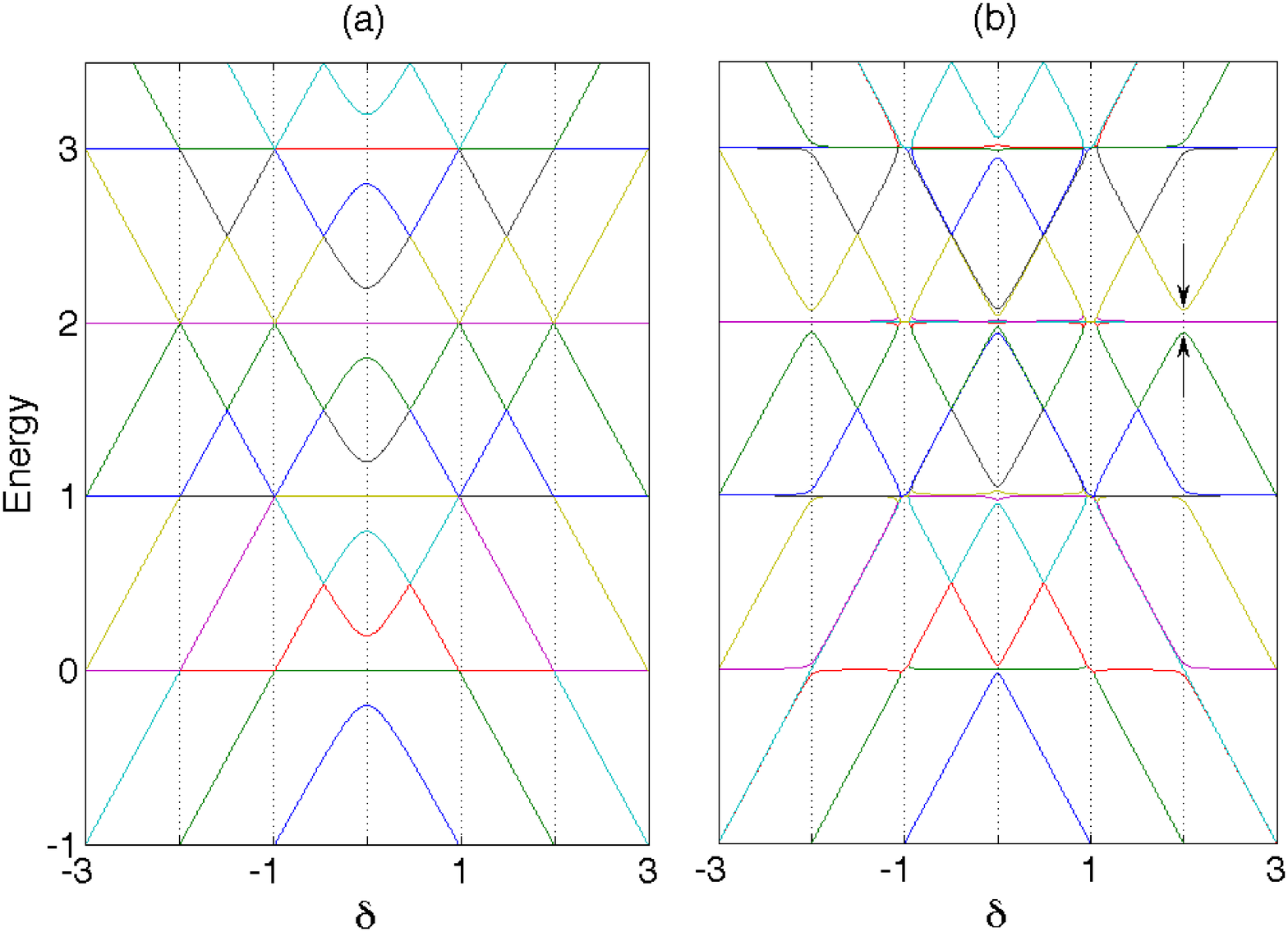}
\caption{Energy spectra for two non-interacting two-level Bose particles. The Rabi frequency $\Omega=0.2$. Here, we only consider three lowest vibrational states (i.e. $l=\{0, 1, 2\}$). The SOC strength is chosen as (a) $z_0=0$ and (b) $z_0=1.5$.} \label{fig. 8}
\end{figure}
%%%%%%%%%%%%%%%%%%%%%%%%%%%%%%%%%%%%%%%%%%%%%%%%%%%%%%%%%%%%%%%%

In Fig.~9, we show the energy spectra for an interacting two-particle system. By adjusting interaction strengthes, we explore the interplay between the interaction blockade and the FC blockade. For simplicity, we assume the intra-component interactions $g=g_{\downarrow\downarrow}=g_{\uparrow\uparrow} > 0$ (repulsive). The inter-component interaction $g_{\downarrow\uparrow}$ is changed from negative (attractive) to positive (repulsive). In our calculation, the parameters are chosen as $\hbar=1$, $\omega=1$, $\Omega=0.04$, $z_0=0.8$ and $g = 0.1$.

If the inter-component interaction $g_{\downarrow\uparrow} <0$ and it is sufficiently strong, the interaction blockade dominates the LZ process from the ground state $\left|2_{\downarrow}^{0}\right\rangle$, see Fig.~9~(a). For a sufficiently slow sweeping process, the system undergoes adiabatic evolution along the path, $\left|2_{\downarrow}^{0}\right\rangle \Longrightarrow \left|1_{\downarrow}^{0},1_{\uparrow}^{0}\right\rangle \Longrightarrow \left|2_{\uparrow}^{0}\right\rangle$, which is marked by three arrows. This evolution path means that the atoms change their spins one by one, which is just the resonant single-atom tunneling induced by the interaction blockade~\cite{Capelle, Kivshar, Cheinet}. In this LZ process, the sequential LZ transitions are dominated by the interaction blockade and the signatures of FC blockade disappear.

Increasing the inter-component interaction strength $g_{\downarrow\uparrow}$ to $0.4$, the structure of the first avoided ELC is almost not changed by the inter-particle interactions, see Fig.~9~(b). Similar to the non-interacting case shown in Fig.~8, this avoided ELC involves the three states: $\left|2_{\downarrow}^{0}\right\rangle$, $\left|1_{\downarrow}^{0}, 1_{\uparrow}^{0}\right\rangle$ and $\left|2_{\downarrow}^{0}\right\rangle$. The unchanged ELC structure is a results of the balance between inter- and intra-component interactions. For a system without SOC (i.e. $z_0=0$), the interaction balance occurs at $g_{\downarrow\uparrow}=g$. However, in our system with the SOC strength $z_0=0.8$, it is obviously that $g_{\downarrow\uparrow}=0.4$ is not equal to $g=0.1$. Actually, because the effective interactions $U_{\sigma\sigma'}^{l_{1}l_{2}l_{3}l_{4}}$ are the products of the interaction strengthes $g_{\sigma\sigma'}$ and the overlap integrals $Y^{l_{1}l_{2}l_{3}l_{4}}$, the interaction balance at $g_{\downarrow\uparrow} \ne g$ originates from the decrease of the inter-component overlap integrals induced by the FC effects. In the LZ process from $\left|2_{\downarrow}^{0}\right\rangle$ with a suitable sweeping rate, as marked by the arrows, the system may adiabatically evolve into the final state $\left|2_{\uparrow}^{0}\right\rangle$.

%%%%%%%%%%%%%%%%%%%%%%%%%%%%%%%%%%%%%%%%%%%%%%%%%%%%%%%%%%%%%%%%
\begin{figure}[htb]
\includegraphics[width=1.0\columnwidth]{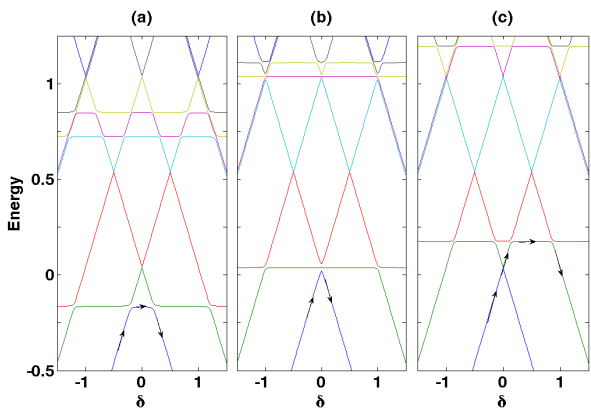}
\caption{Energy spectra for two interacting two-level Bose atoms with spin-orbit coupling. The interplay between FC blockade and interaction blockade is explored by adjusting the inter-component interaction strength: (a) $g_{\downarrow\uparrow}=-1$, (b) $g_{\downarrow\uparrow}=0.4$ and (c) $g_{\downarrow\uparrow}=5$. The other parameters are chosen as $\hbar=1$, $\omega=1$, $z_{0}=0.8$, $\Omega=0.04$ and $g=g_{\downarrow\downarrow}=g_{\uparrow\uparrow}=0.1$. The values of $U_{\sigma\sigma'}^{l_{1}l_{2}l_{3}l_{4}}$ in the multi-mode two-component Bose-Hubbard Hamiltonian are given by the product of interaction strengthes $g_{\sigma\sigma'}$ and the overlap integrals $Y^{l_{1}l_{2}l_{3}l_{4}}$.}
\label{fig. 9}
\end{figure}
%%%%%%%%%%%%%%%%%%%%%%%%%%%%%%%%%%%%%%%%%%%%%%%%%%%%%%%%%%%%%%%%

Increasing the inter-component interaction strength $g_{\downarrow\uparrow}$ to $5$, the energy level for $\left|1_{\downarrow}^{0}, 1_{\uparrow}^{0}\right\rangle$ is lifted from the first ELC and the energy gap between $\left|2_{\downarrow}^{0}\right\rangle$ and $\left|2_{\uparrow}^{0}\right\rangle$ is closed, see Fig.~9~(c). Therefore, the population transition between $\left|2_{\downarrow}^{0}\right\rangle$ and $\left|2_{\uparrow}^{0}\right\rangle$ becomes ignorable. In the LZ process from $\left|2_{\downarrow}^{0}\right\rangle$, by selecting a suitable sweeping rate, the system will evolve along the path $\left|2_{\downarrow}^{0}\right\rangle \Longrightarrow \left|1_{\downarrow}^{0}, 1_{\uparrow}^{0}\right\rangle \Longrightarrow \left|1_{\uparrow}^{0},1_{\uparrow}^{1}\right\rangle$ which is marked by four arrows. In this LZ process, the atoms change their spins one by one, which is a manifestation of the interaction blockade. However, at the same time, the vibrational state of one of two particles changes from the lowest vibrational state $\left|l=0\right\rangle$ to the first-excited vibrational state $\left|l=1\right\rangle$. This is different from the LZ process marked in Fig.~9~(a), in which the vibrational states of all atoms keep unchanged. The simultaneous changes of spin and vibrational states are a direct result of the cooperation between the SOC effects and the interaction blockade.

\section{Summary and discussion}

In summary, we have studied the LZ process of two-level particles in a spin-dependent harmonic trap. The spin-orbit coupling, the coupling between spin and vibrational states, is induced by the spin-dependent harmonic potential. We consider the single-particle systems at first and explore the intrinsic mechanism of the sequential LZ transitions induced by the SOC. Then, we consider the multi-particle system and explore the interplay between the SOC and the inter-particle interactions. The sequential LZ transitions provide a new perspective for exploring signatures of FC effects and interaction blockade. Further, our results may also provide potential applications in quantum state engineering.

Considering the spin-orbit coupled single-particle system, the intrinsic mechanism of the sequential LZ transitions and the direct signatures of the FC effects in the LZ process have been explored. The sequential LZ transitions, which have ladder-like population transitions, take place when the SOC is sufficiently strong. The presence of population steps at $\delta =n\hbar \omega _{z}$ is a signature of the vibrational sidebands and the absence of some specific steps is a signature of the FC blockade.

The single-particle population dynamics depends upon both the SOC strength and the sweeping rate. However, interestingly, the final spin populations are independent upon the SOC strength. In other words, the spin inversion efficiency is only determined by the sweeping rate although the population dynamics in the LZ process are determined by both the SOC strength and the sweeping rate. Further, by selecting suitable sequential LZ transitions, it is possible to implement quantum state engineering and prepare the desired entanglement between spin and vibrational states.

To obverse the ladder-like population steps in the sequential LZ transitions in the single-particle systems, one has to select a moderate sweeping rate. For a small weeping rate, there is only one populations step similar to the LZ process without SOC. For a large sweeping rate, the non-adiabatic effects submerge the ladder-like population steps in the sequential LZ process. The sequential LZ transitions can be treated as a sequence of conventional two-level LZ transitions. We derive an analytical formula for the final populations.

Beyond the single-particle systems, we consider the interacting multi-particle systems and explore the interplay between the SOC effects and the interaction effects. For simplicity, we study the coupled two-component atomic BEC in a spin-dependent harmonic trap, which is described by a multi-mode two-component Bose-Hubbard Hamiltonian. In addition to the FC blockade, the interaction blockade appears when the system is dominated by the interaction effects. In a pure interaction blockade, the atoms change their spin states one by one and their vibrational states keep unchanged. It is also possible to find the cooperation between the SOC effects and the interaction effects, in which the atoms change simultaneously their spin and vibrational states.

It is possible to realize our spin-orbit coupled systems by current experimental techniques. By using a suspended carbon nanotube quantum dot~\cite{Palyi}, a single ion in spin-dependent Paul trap~\cite{Hu}, or ultracold atoms in a spin-dependent optical lattice~\cite{Forster}, it is possible to test our prediction in experiments. Here, we briefly discuss the cases of a single trapped ion~\cite{Hu} or ultracold atoms~\cite{Forster}. In the case of a single trapped ion, the spin-dependent potential can be formed by imposing a gradient magnetic field on the Paul trap~\cite{Mintert} and the SOC strength is determined by the field gradient. In the case of ultracold atoms, the spin-dependent optical lattice~\cite{Mandel} can be created by two polarized counter-propagating lasers and the SOC strength can be adjusted by tuning the polarization angle. In both two cases, the two internal spin states can be coupled by Raman lasers and the detuning can be varied by modifying the laser frequency.

\acknowledgements

This work is supported by the National Basic Research Program of China under Grant No. 2012CB821300, the National Natural Science Foundation of China under Grants No. 11075223 and No. 10804132, the Program for New Century Excellent Talents in University of Ministry of Education of China under Grant No. NCET-10-0850 and the Ph.D. Programs Foundation of Ministry of Education of China under Grant No. 20120171110022.

\end{document}